\documentclass{aa}  

\usepackage{txfonts}
\usepackage{color}
\usepackage{comment}
\usepackage{orcidlink}
\usepackage{natbib}
\usepackage{url}
\usepackage{hyperref}
\usepackage{subfigure}
\usepackage{xspace}
\usepackage[percent]{overpic}
\usepackage{subfloat}

\usepackage{amsmath,amstext}
\usepackage{multirow}
\usepackage{hyperref}

\newcommand{\ixpe}{{\it IXPE}\xspace}

\newcommand{\nustar}{{\it NuSTAR}\xspace}

\newcommand{\xmm}{{\it XMM-Newton}\xspace}

\newcommand{\src}{SRGA~J1444}

\graphicspath{{./}{figures/}}

\begin{document} 
\title{Disk reflection and energetics from the accreting millisecond pulsar SRGA~J144459.2$-$604207}
\titlerunning{\src\ reflection and energetics}
\authorrunning{Malacaria et al.}

\author{
Christian~Malacaria\orcidlink{0000-0002-0380-0041} \inst{\ref{in:INAF-OAR}}
\and 
Alessandro~Papitto \inst{\ref{in:INAF-OAR}}\orcidlink{0000-0001-6289-7413}
\and
Sergio Campana\inst{\ref{in:INAF-MI}}\orcidlink{0000-0001-6278-1576}
\and
Alessandro Di Marco\orcidlink{0000-0003-0331-3259}\inst{\ref{in:INAF-IAPS}}
\and
Tiziana~Di~Salvo\inst{\ref{in:UniPa}}\orcidlink{0000-0002-3220-6375}
\and
Maria~Cristina~Baglio\inst{\ref{in:INAF-MI}}\orcidlink{0000-0003-1285-4057}
\and
Giulia Illiano\inst{\ref{in:INAF-OAR},\ref{in:INAF-MI}}\orcidlink{0000-0003-4795-7072}
\and 
Riccardo La Placa\inst{\ref{in:INAF-OAR}}\orcidlink{0000-0003-2810-2394}
\and
Arianna~Miraval~Zanon\inst{\ref{in:ASI}}\orcidlink{0000-0002-0943-4484}
\and
Maura Pilia\inst{\ref{in:INAF-OAC}}\orcidlink{0000-0001-7397-8091}
\and 
Juri~Poutanen \inst{\ref{in:Turku},\ref{in:IKI}}\orcidlink{0000-0002-0983-0049}
\and
Tuomo~Salmi\inst{\ref{in:Helsinki}}\orcidlink{0000-0001-6356-125X}
\and
Andrea Sanna \inst{\ref{in:UniCa}}\orcidlink{0000-0002-0118-2649}
\and
Manoj~Mandal\inst{\ref{in:Indian}, \ref{in:Indian-PRL}}\orcidlink{0000-0002-1894-9084}
}
\institute{
INAF-Osservatorio Astronomico di Roma, Via Frascati 33, 00076 Monte Porzio Catone (RM), Italy \label{in:INAF-OAR} \\ \email{christian.malacaria@inaf.it}
\and INAF-Osservatorio Astronomico di Brera, Via Emilio Bianchi 46, 23807 Merate (LC), Italy \label{in:INAF-MI}
\and INAF Istituto di Astrofisica e Planetologia Spaziali, Via del Fosso del Cavaliere 100, 00133 Roma, Italy \label{in:INAF-IAPS}
\and Dipartimento di Fisica e Chimica – Emilio Segrè, Università di Palermo, Via Archirafi 36, 90123 Palermo, Italy \label{in:UniPa}
\and ASI - Agenzia Spaziale Italiana, Via del Politecnico snc, 00133 Rome (RM), Italy \label{in:ASI}
\and INAF/OAC, via della Scienza 5, I-09047, Selargius (CA), Italy \label{in:INAF-OAC}
\and Department of Physics and Astronomy, 20014 University of Turku, Finland \label{in:Turku}
\and Space Research Institute, Russian Academy of Sciences, Profsoyuznaya 84/32, Moscow 117997, Russia \label{in:IKI}
\and Department of Physics, P.O. Box 64, 00014 University of Helsinki, Finland \label{in:Helsinki}
\and University of Cagliari, SP Monserrato-Sestu km 0.7, 09042 Monserrato, Sardinia, Italy\label{in:UniCa}
\and Midnapore City College, Kuturia, Bhadutala, Paschim Medinipur, West
Bengal, 721129, India \label{in:Indian}
\and Astronomy and Astrophysics Division, Physical Research Laboratory, Navrangpura, Ahmedabad - 380009, Gujarat, India \label{in:Indian-PRL}
}

   \date{\today}

 
  \abstract
   {Accreting millisecond pulsars (AMSPs) are excellent laboratories to study reflection spectra and their features as emission is reflected off an accretion disk truncated by a rapidly rotating magnetosphere near the neutron star  surface. These systems also exhibit thermonuclear (type-I) bursts that can provide insights on the accretion physics and fuel composition.}
   {We explore spectral properties of the AMSP SRGA~J144459.2$-$604207 observed during the outburst that recently led to its discovery in February 2024.
   We aim to characterize the spectral shape of the persistent emission, both its continuum and discrete features, and to analyze type-I bursts properties.}
   {We employ \xmm\ and \nustar\ overlapping observations taken during the most recent outburst  from SRGA~J144459.2$-$604207.
   We perform spectral analysis of the time-averaged persistent (i.e., non-bursting) emission. For this, we first employ a semi-phenomenological continuum model composed of a dominant thermal Comptonization plus two thermal contributions. A separate fit has also been performed employing a physical reflection model. We also perform time-resolved spectral analysis of the type-I bursts employing a blackbody model.}
   {We observe a broadened iron emission line, thus suggesting relativistic effects, supported by the physical model accounting for relativistically blurred reflection. The resulting accretion disk extends down to 6 gravitational radii, it is observed at an inclination of $\sim53\degr$, and is only moderately ionized ($\log\xi\simeq2.3$). We observe an absorption edge at $\sim9.7\,$keV that can be interpreted as an \ion{Fe}{xxvi} edge blueshifted by an ultrafast ($\simeq0.04c$) outflow. Our observations of type-I bursts also allowed us to characterize the broadband emission evolution during the burst. We do not find evidence of photospheric radius expansion. Regarding the burst recurrence time we observe a dependence on the count rate that has the steepest slope ever observed in these systems. We also observe a discrepancy by a factor $\sim3$ between the observed burst recurrence time and its theoretical expected value, which we discuss in the framework of fuel composition and high NS mass scenarios.}
   {}
  

   \maketitle
%

\section{Introduction}

Accreting millisecond pulsars (AMSPs) are neutron stars (NSs) rotating with a spin frequency of >100 Hz found in tight binary systems (orbital periods $P_{\rm orb}\sim$~hours) with low-mass companion stars (see, e.g., \citealt{Patruno+Watts2021, DiSalvo+Sanna2022} for recent reviews). 
AMSPs show X-ray outbursts driven by unsteady accretion of matter when the X-ray flux increases by orders of magnitude. 
The source SRGA J144459.2$-$604207 (SRGA~J1444 hereafter) is the most recently discovered AMSP at the time of writing. 
First discovered as a new X-ray source \citep{Mereminskiy2024} with the \textit{Mikhail Pavlinsky} {ART-XC} telescope \citep{Pavlinsky21}   on board of the {Spectrum-Roentgen-Gamma observatory} \citep{Sunyaev21}, the source was then identified as a 447.9~Hz AMSP  with {NICER}  in a binary system with the orbital period of 5.2~hr and a 0.2--0.7\,${M_\odot}$ donor companion \citep{Ng2024}.
Analysis of the {ART-XC} also returned a wealth of information about spectral and energetic characteristics of the source, including the estimation of the distance as 8--9 kpc \citep{Molkov2024}.
No optical or near-infrared counterparts are known within the accurate \textit{Chandra} localization \citep{Illiano2024ATel}.
\src\ was also recently reported as the first AMSP to show polarized emission at an average polarization degree of $2.3\%\pm0.4\%$ as observed by the Imaging X-ray Polarimetry Explorer  \citep[\ixpe;][]{Papitto2025}.
The energy spectrum from an AMSP is typically described by a Comptonized component with an electron temperature of tens of keV plus a relatively hot (${\sim}$0.5--1\,keV) blackbody component interpreted as thermal emission from the NS hotspot or from the boundary layer, to which sometimes a colder blackbody component representing emission from the accretion disk is added (see, e.g., \citealt{Gierlinski2002, Falanga2012, Illiano2024}).
A reflection component may also arise from the accretion disk illumination, showing up in the spectrum as an iron K$\alpha$ complex around 6.4 keV and a hump in the hard X-ray band \citep{Fabian1989, Barret2000, DiSalvo2015}.
When the iron line is present, its shape shows signs of relativistic broadening, with an asymmetric profile covering the range up to 4--8\,keV \citep[and references therein]{Illiano2024}, although alternative explanations for the unusual profile have been also proposed (see, e.g., \citealt{Disalvo2009} for a discussion of relevant works). 
During outbursting episodes, recurrent type-I bursts are observed, interpreted as unstable thermonuclear burning flashes ignited at the magnetic polar caps on the NS surface \citep{Strohmayer2003, Galloway2008}. 

In this work we analyze the broadband spectral properties of the newly discovered AMSP \src\ observed with \xmm\ and \nustar. We adopt semi-phenomenological and physically motivated models to describe the source spectrum in order to infer physical properties of the system. We report on the discovery of a relativistically broadened iron line in this source. We also analyze the spectral evolution over one of the several type-I bursts observed from the source and report on the burst energetics.


\section{Observations}
\subsection{\nustar\ observations}

\nustar\ \citep{Harrison2013} observed \src\ starting from 2024-02-26T11:01:09 for about 157 ks (ObsID 80901307002).
Source events were extracted from circular regions centered on the source with an extraction radius of 100\arcsec, while background events were extracted from an empty circular region with a similar radius on the same detector where the source lies.
Source spectra were rebinned with the \texttt{ftgrouppha} tool with the ``optimal'' grouping option following the \citet{Kaastra2016} binning algorithm plus minimum of 40 counts per grouped bin, while background spectra have been rebinned to contain at least 3 counts per bin in order to avoid background biases (similarly to, e.g., \citealt{Snios2020}).
\nustar\ CALDB version 20240729 was used. The final exposure is about 124 ks for each FPM.

\subsection{\xmm\ observations}\label{subsec:xmm_analysis}

\xmm\ observed \src\ starting from 2024-02-28T02:26:21 for about 128 ks (ObsID 0923171501) with EPIC-pn in the Timing mode (30 $\mu$s resolution) and with the Reflection
Grating Spectrograph (RGS).
\xmm\ data were reduced using the \xmm\ Science Analysis System (SAS) v21.0.0 and the latest available calibration files (XMM-CCF-REL-403). The standard procedure described in the SAS Data Analysis Threads was followed for the data reduction,\footnote{\url{https://www.cosmos.esa.int/web/XMM-Newton/sas-threads.}} resulting in a final exposure of about 107 ks.

Given the timing observation mode, the source is visible on the EPIC-pn detector as a bright strip instead of the usual circular shape. After the standard cleaning of raw data, events were extracted from a 20-pixel wide box-shaped region centered at the source. Background events were extracted from a 3-pixel width box far from the source. 
After obtaining a light curve, times around the bursts activity (that is, between 50~s before and 100s after the peak) were excluded from the GTIs to generate a spectrum of the persistent emission. 
Spectral response files were generated using the tasks \texttt{rmfgen} and \texttt{arfgen}.
Although the observation count rate ($\simeq150\,$cnt\,s$^{-1}$) is well below the nominal limit for pile-up contamination (i.e., 800\,cnt\,s$^{-1}$), we ran tests for the presence of pile-up. We compared the \texttt{epatplot} verification plot and found a non-negligible pile-up. We therefore compared pile-up effects obtaining verification plots for events in which one, three, five, and seven central pixel were excised. The most effective pile-up reduction was obtained by excising the three central pixels, and therefore we adjusted our cleaning procedure to reflect that.

\begin{figure}
    \centering
    \includegraphics[width=0.65\columnwidth, height=\columnwidth, angle=-90]{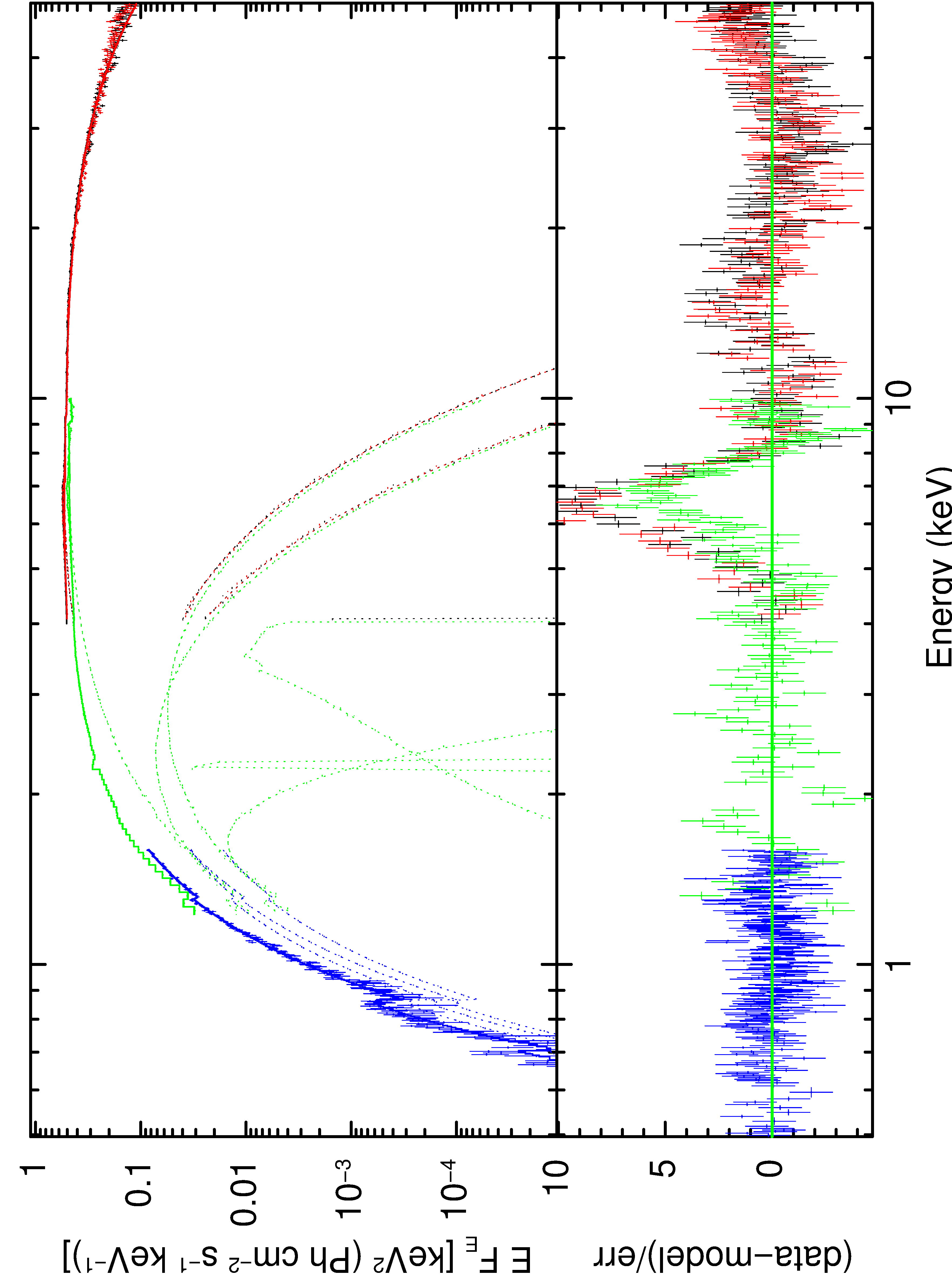}
    \caption{Unfolded average spectrum (0.5--50\,keV) of the persistent emission from \src. Top panel: merged \xmm\ RGS (blue points), EPIC-pn (green points) and \nustar\ FPMA and FPMB (black and red points, respectively) are shown as fitted to the semi-phenomenological best-fit model 
    \texttt{constant*tbabs(bbodyrad + diskbb + nthComp + gau$_{\rm Si}$+ gau$_{\rm Au}$+ diskline$_{\rm Ar}$ + diskline$_{\rm Fe}$)*edge*edge*edge} (see Sect~\ref{subsection:iron_line}).
    Bottom panel: residuals in units of $\chi$ with respect to the best-fit model but with the \texttt{diskline$_{\rm Fe}$} component at 6.4 keV removed.}
    \label{fig:iron_line}
\end{figure}

RGS data were processed using the \texttt{rgsproc} pipeline, obtaining first and second order source and background spectra, and response matrices. 
After careful checking that the selection regions are centered on the prime source, we excluded the background light curve time displaying particle flaring events, and obtained new GTIs for a suitable data reprocessing. RGS1 and RGS2 data were merged using the tool \texttt{rgscombine}, keeping separate the first and second order spectra. However, in the following only the first order resulting spectrum will be considered.

EPIC-pn and RGS spectra were rebinned using the SAS \texttt{specgroup} tool in order to have at least 25 counts per bin and not to oversample the intrinsic instrumental energy resolution by a factor larger than three.

\section{Spectral analysis of the persistent emission}\label{sec:persistent_emission}

Spectral data were analyzed using \textsc{xspec} {v12.14.0} \citep{Arnaud96} as part of \textsc{heasoft} v6.33.1.
To test the origin of the X-ray persistent emission from \src, we selected GTIs such that type-I bursts are ignored at this stage (see Sect.~\ref{sec:time-resolved} for more details). 
In all tested models, the photoelectric absorption component (with column density $N_{\rm H}$) was modeled according to \citet[\texttt{tbabs} in \textsc{xspec}]{Wilms00} 
and we assumed \texttt{wilm} Solar abundances.
To derive best-fit spectral parameters we employed the Cash statistics \citep{Cash79} with Poissonian background (W-stat) as a fit statistic while we used the $\chi^2$ test statistic to evaluate the best-fit model.
Spectral fits were performed using broadband data in the 0.5--50 keV range.
In the following, we describe the fitting procedure first using a phenomenological model  (Sect.~\ref{subsection:iron_line}) and then with a physical reflection model  (Sect.~\ref{subsection:reflection}).

\subsection{Semi-phenomenological spectral modeling}\label{subsection:iron_line}

Relativistically broadened iron emission lines are observed in AMSPs \citep[and references therein]{Illiano2024}. To investigate such a feature in \src\ we adopt a semi-phenomenological continuum spectral model consisting of an absorbed Comptonized component modified by 
additional components in the softer energy domain \citep{Disalvo2009, Papitto2009,Papitto2010,DiSalvo2019}.
The employed Comptonized component is \texttt{nthComp} with photon index, electron temperature, seed photon temperature, and normalization as free parameters \citep{Zdziarski1996, Zycki1999}.
Two thermal components are required to account for the soft spectrum, a hotter \texttt{bbodyrad} component, typically interpreted as originating from the NS surface, and a colder \texttt{diskbb} component, typically interpreted as emission from the inner regions of the accretion disk.
Moreover, residual emission features are evident along the soft energy coverage of \xmm. Although some of those features might be of astrophysical origin, some others are likely instrumental. The Si-K line at 1.84~keV and the Au-M line at 2.2~keV are reported by the \xmm\ team as known artificial features due to instrumental edges that are not perfectly calibrated (see the official \xmm\ user guide\footnote{\url{https://xmm-tools.cosmos.esa.int/external/xmm_user_support/documentation/sas_usg/USG/epicdataquality3.html}}\footnote{\url{https://www.cosmos.esa.int/web/xmm-newton/sas-thread-evenergyshift}}). We modeled these components with a \texttt{gauss} profile.
Besides, we find evidence of additional features: one emission line with centroid energy of 3.3 keV and three edges at 0.82, 1.32, and 9.68 keV.
Several of these features are also observed in other RGS and EPIC-pn observations and are, at least in part, likely of instrumental origin. Following \citet{Pintore16} we identify the emission line at 3.3 keV with \ion{Ar}{xviii}, and the first two edge features with Ne and Mg, respectively.
Alternatively, these can be identified as redshifted OVIII (0.871 keV) e NeX (1.362 keV) edges, or as blushifted OVII (0.739 keV) e Ne IX (1.196 keV).
The edge at 9.68 keV is consistent with an iron K-edge of high ionization state \citep{Makishima1986} or, alternatively, with a Zn edge (nominally at 9.65 keV).
We modeled these edges with  \texttt{edge} profile.

Joint spectral fits of EPIC-pn spectra obtained in timing mode with \nustar\ spectra are known to exhibit inconsistencies probably due to calibration uncertainties \citep{DiSalvo2019, Diez2023}.
To mitigate those we also employed a cross-normalization constant among all detectors, which resulted about $16\%$ off between EPIC-pn and the FPMs (consistently with recent calibration results, see e.g., \citealt{Saavedra2023} and the 2022 IACHEC report\footnote{\url{https://iachec.org/wp-content/presentations/2022/Fuerst_pn_NUSTAR.pdf}}).
Following the latest \xmm\ calibration results\footnote{G. Matzeu 2024, \url{https://www.cosmos.esa.int/web/xmm-newton/epic-calibration-meetings\#202406}} we also tested that the photon index $\Gamma$ for the EPIC-pn spectrum, when left independent from \nustar\ during the joint fit, results in a difference of less than $10\%$, which is well within the expected values.

\begin{figure}[!t]
    \centering
\includegraphics[width=0.65\columnwidth, height=\columnwidth, angle=-90]{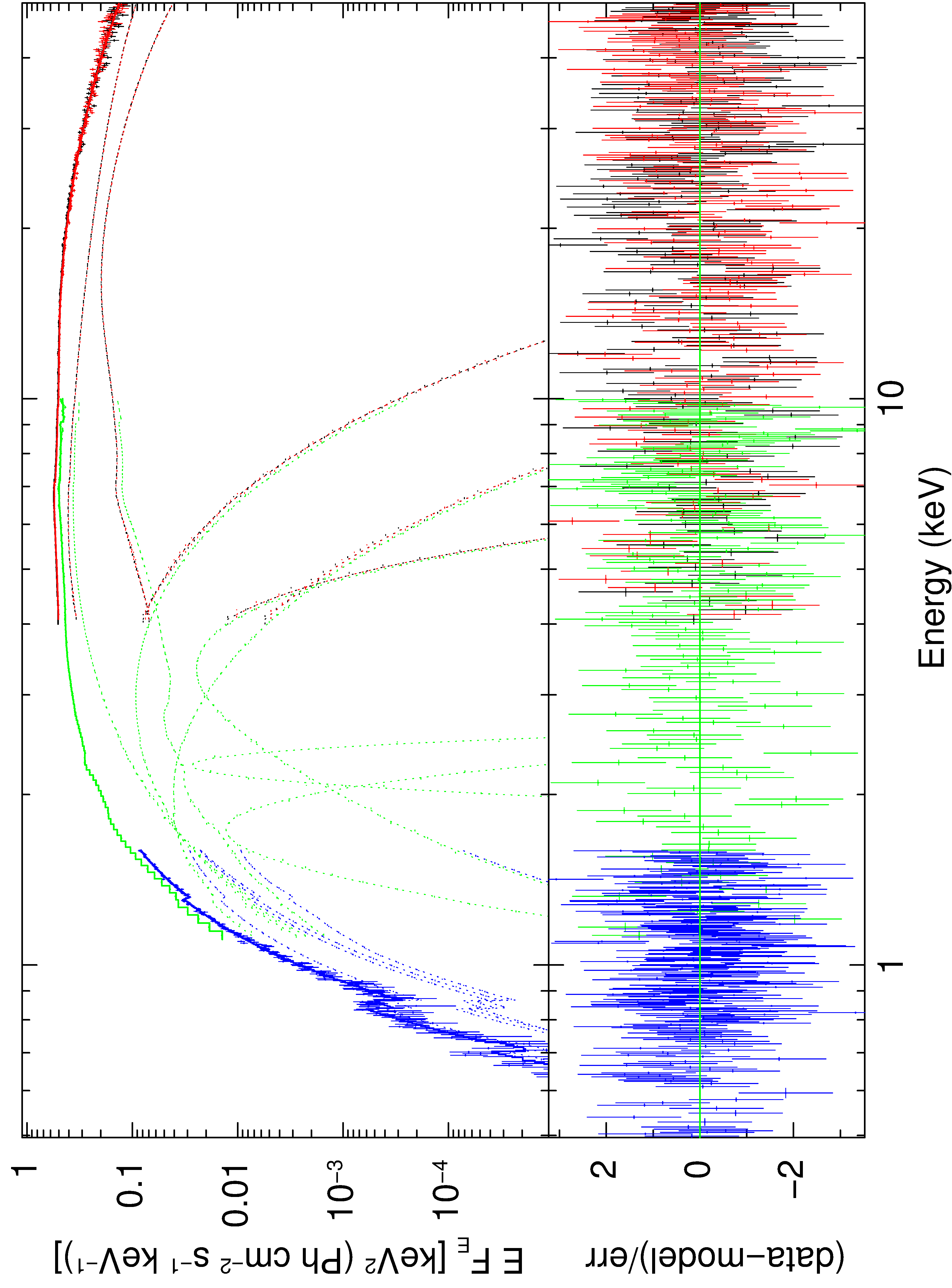}
    \caption{Similar to Fig.~\ref{fig:iron_line}. Top panel: data and spectra for the best-fit continuum model composed of \texttt{constant*tbabs(bbodyrad + diskbb + nthComp + gau$_{\rm Si}$+ gau$_{\rm Au}$+ rdblur*gau$_{\rm Ar}$ + rdblur*rfxconv*nthComp)*edge*edge *edge} (see Table\ref{table:spectral_xmm_nustar}). Bottom panel: residuals in units of $\chi$.}
    \label{fig:persistent}
\end{figure}

Adopting the components of the spectral model mentioned above, clear residuals appear around the iron K$\alpha$ energy at 6.4~keV, as indicated by the broad residuals in Fig.~\ref{fig:iron_line}.
We model this feature with a \texttt{diskline} component.
The parameters of the \texttt{diskline} component are the emissivity index \textit{Betor}, that is the power-law index of the illuminated disk emissivity (scaling as $r^{Betor}$), inner and outer disk radius, and disk inclination \citep{Fabian1989}.
Moreover, as the Ar line is broad and likely originating in the disk (thus participating of the same relativistic effect that shape the iron line), we also model the Ar line using a \texttt{diskline} component whose inclination and inner disk radius were linked to those of the iron line.
When fitting the iron \texttt{diskline} component, if all line parameters are let free then the best-fit line energy is 
$6.2^{+0.1}_{-0.1}$\,keV (at $1\sigma$\,c.l.). 
The remaining best-fit parameters of the line are the \textit{Betor} parameter, $-2.6^{+0.1}_{-0.1}\,$, inner disk radius $10.0^{+1.2}_{-1.7}\,\mathrm{R}_{g}$ (where $\mathrm{R}_{g}=GM_{\rm NS}/c^2$ is the gravitational radius), and the inclination $85^{+5}_{-14}$\,deg.

Alternatively, to retrieve an iron line energy that falls in the K$\alpha$ iron energy range (i.e., 6.40--6.97\,keV) we freeze the inclination parameter of the \texttt{diskline} components to the value obtained by fitting the reflection spectrum, that is 53\degr\ (see Sect.~\ref{subsection:reflection}).
The remaining best-fit parameters of the iron line model component are the energy of the line, $6.41^{+0.03}_{-0.03}\,$keV, the \textit{Betor} parameter, $-2.45^{+0.05}_{-0.05}\,$, and the inner disk radius, $6.0^{+0.3}_{}\,R_{\rm g}$, where the lower value is pegged to the model limit of $6\,R_{\rm g}$ (see, e.g., \citealt{Cackett2010}).
Employing this iron \texttt{diskline} component brings the test statistic  to $\chi^2$/d.o.f. = 1329/722 (where d.o.f. refers to the degrees of freedom) with respect to a fit statistic of $\chi^2$/d.o.f. = 1735/724 where the iron \texttt{diskline} component was not taken into account in the model.

\subsection{Modeling reflection}\label{subsection:reflection}

\begin{figure}[!t]
    \centering
\includegraphics[height=0.9\columnwidth, angle=-90]{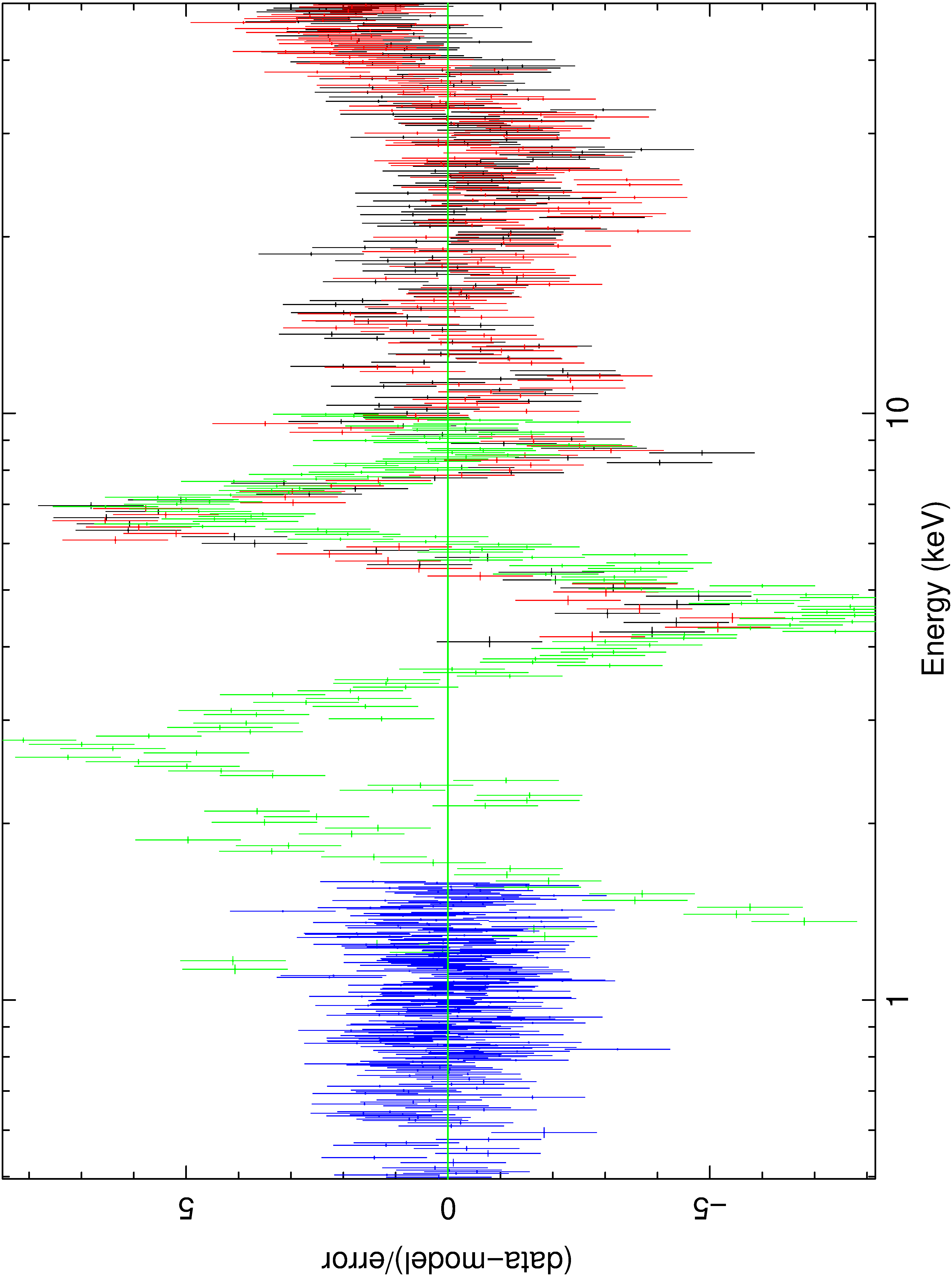}
    \caption{Residuals of the model shown in Fig.~\ref{fig:persistent} with the reflection component removed. Several spectral features are evident, including the iron line at 6.4 keV (see text).}
    \label{fig:residuals}
\end{figure}

We fit \xmm\ and \nustar\ spectral data with a reflection model modified by several components.
The dominant components are those described in Sect.~\ref{subsection:iron_line} to which, motivated by the presence of a broadened iron line, we added a reflection component.
We account for the reflected component using the convolution model \texttt{rfxconv} \citep{Done2006} with the relativistic blurring kernel \texttt{rdblur}.
The \texttt{rfxconv} model parameters are the relative normalization of the reflection component, the iron abundance (relative to solar ones), the cosine of the disk inclination angle and the ionization parameter, while the \texttt{rdblur} parameters are, similarly to the \texttt{diskline} component described in Sect.~\ref{subsection:iron_line}, the parameter \textit{Betor}, the inner and outer disk radii, and the disk inclination.
We therefore modeled the \ion{Ar}{xviii} line at 3.3 keV with a \texttt{rdblur*gauss} model, which is equivalent to the \texttt{diskline} component, linking the \texttt{rdblur} parameters to those of the reflection component.
The inclusion of this component implies a further reduction with respect to the semi-phenomenological model to $\chi^2$/d.o.f. = 881/719.
A visualization of the reflection features accounted by this component is shown in Fig.~\ref{fig:residuals}.
Based on the F-test, the probability of chance improvement due to the inclusion of the \texttt{rfxconv} component is $10^{-63}$.
The resulting best-fit model is thus \texttt{constant*tbabs(bbodyrad + diskbb + nthComp + gau$_{\rm Si}$+ gau$_{\rm Au}$+ rdblur*gau$_{\rm Ar}$ + rdblur*rfxconv*nthComp)*edge*edge*edge} in \textsc{xspec} terminology.

As a sanity check, we also tested the spectral fit results by ignoring the data in the iron K$\alpha$ energy band (4.5--8.0\,keV) and re-fitting the data. The resulting iron abundance remains low (of the order of 20\%), although the high-energy hump due to reflection remains important. This is in agreement with results from \citet{Garcia2022} of X-ray reflection spectra explored at different iron abundances, which show that the high-energy hump is almost insensitive to the iron abundance, especially for low abundances and low ionization parameter values.
For completeness, we also checked the alternative procedure in which we freeze the iron abundance to the Solar one and repeated the fitting procedure excluding spectral data in the iron K$\alpha$ energy band.
This fit gives $\chi^2$/d.o.f.=769/623, but the Ar line around 3.3 keV shows a width of $\sigma\sim0.8\,$keV, too large to be an atomic spectral feature. Also, the $Betor$ parameter is pegged at its minimum value of $-10$. If data around the iron line energy are included back then $\chi^2=9234$, with significant negative residuals showing up, indicating that the model over-predicts the line intensity due to the fixed abundance and the high (absolute) value of $Betor$, which makes the emissivity of the inner disk drastically growing. Upon re-fitting the spectra we obtain $\chi^2$/d.o.f.=1038/720, and the reflected fraction drops to $-0.2$, while the parameter $Betor$ settles at $-2.5$. We therefore adopted the best-fit model described above.
Our best-fit spectral analysis results are reported in Table~\ref{table:spectral_xmm_nustar}.

\begin{table}[!t]
\caption{Best-fit parameters of \src\ data from \xmm\ (RGS and PN) +\nustar\ observations of the persistent emission fit jointly. All errors are reported at $1\sigma\,$c.l. The model is \texttt{constant*tbabs(bbodyrad + diskbb + nthComp + gau$_{\rm Si}$+ gau$_{\rm Au}$+ rdblur$^*$*gau$_{\rm Ar}$ + rdblur*rfxconv*nthComp)*edge*edge *edge}.} \label{table:spectral_xmm_nustar}
\centering
\begin{tabular}{lll}
\hline\hline  
Model & Parameter & Value \\
\hline 
\texttt{tbabs} & $N_{\rm H}$ [$10^{22}\,$cm$^{-2}$] & $1.92^{+0.03}_{-0.01}$\\
\texttt{bbodyrad} & temp [keV] & $0.64^{+0.02}_{-0.01}$\\
         & Radius [norm] & $179^{+103}_{-64}$ \\
\texttt{diskbb} & Temp [keV] & $0.44^{+0.09}_{-0.06}$\\
         & Radius [norm] & $664^{+451}_{-292}$ \\
nthComp & $\Gamma$ & $2.39^{+0.01}_{-0.04}$ \\
        & $kT_{\rm e}$ [keV] & $18.2^{+0.7}_{-2.7}$\\
        & $kT_{\rm seed}$ [keV] & $0.93^{+0.02}_{-0.02}$\\
        &norm & $0.039^{+0.002}_{-0.001}$\\
Reflection  &  $Betor$ & $-2.49^{+0.08}_{-0.09}$ \\
            & $R_{\rm in}$ [$R_{\rm g}$] & $6.0^{+1.6}$ \\
            & $R_{\rm out}$ [$R_{\rm g}$] & $1000$ (fix)\\
            & Inclinat. [deg] & $52.7^{+1.9}_{-1.6}$ \\
            & Refl. frac. & $-0.73^{+0.17}_{-0.10}$ \\
            & Iron abund.  & $0.16^{+0.01}_{-0.01}$\\
            & log $\xi$ & $2.31^{+0.05}_{-0.03}$ \\
Edges        & &  \\
Ne 0.87 keV & Energy [keV]& $0.84^{+0.01}_{-0.01}$ \\
            & $\tau$   & $0.69^{+0.07}_{-0.04}$ \\
Mg 1.31 keV & Energy [keV]& $1.32^{+0.01}_{-0.01}$ \\
            & $\tau$   & $0.082^{+0.01}_{-0.01}$ \\
\ion{Fe}{xxvi}? (Zn 9.65) keV & Energy [keV]$^{**}$& $9.68^{+0.13}_{-0.13}$\\
            & $\tau$   & $0.014^{+0.002}_{-0.003}$ \\   
Gaussian emission lines & &  \\
Si K-edge (1.8 keV) & Energy [keV] & $1.83^{+0.01}_{-0.01}$ \\
            & $\sigma$ [keV]  & $0.14^{+0.01}_{-0.01}$ \\
            & norm [$10^{-3}$]  & $5.36^{+0.9}_{-0.8}$ \\
Au M-edge (2.2 keV)& Energy [keV] & $2.24^{+0.01}_{-0.01}$ \\
            & $\sigma$ [keV]  & $0.067^{+0.02}_{-0.01}$ \\
            & norm [$10^{-3}$]  & $1.87^{+0.3}_{-0.2}$ \\
\ion{Ar}{xviii} (3.32 keV) &Energy [keV]& $3.29^{+0.11}_{-0.15}$ \\
            & $\sigma$ [keV]  & $0.40^{+0.06}_{-0.07}$\\
            & norm [$10^{-3}$]  & $5.0^{+2.2}_{-1.3}$ \\
\\
            & Flux$^\dagger$  & $3.225^{+0.005}_{-0.005}$\\
            & $\chi^2$/d.o.f. & 881/719\\
\hline 
\end{tabular}
\tablefoot{$^*$Parameters for this \texttt{rdblur} component are linked to the ones obtained from the reflection component (see text).\quad$^{**}$For this parameter, we also calculated errors at 90\% c.l., resulting in $9.68^{+0.23}_{-0.22}\,$keV (see text).\quad$^\dagger$Unabsorbed flux calculated in the 0.01--100\,keV band and reported in units of $10^{-9}\,$erg\,cm$^{-2}\,$s$^{-1}$.\quad}
\end{table}

\begin{figure*} 
    \centering
    \includegraphics[width=\textwidth]{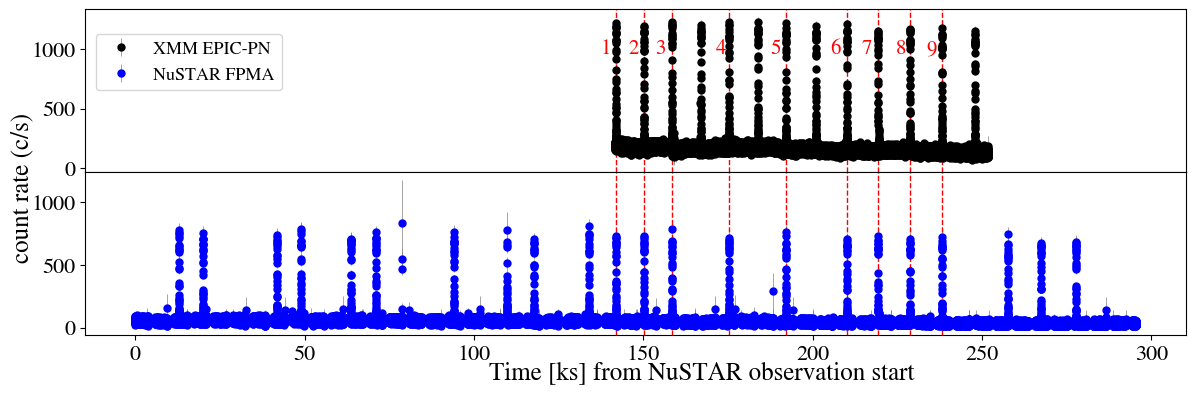}
    \caption{Light curve of \src\ as observed by \nustar\ (bottom) and \xmm\ PN (top) at a time binning of 10\,s.
    Contemporaneously observed type-I bursts are marked with a red vertical dashed line and numbered in the top panel for clarity.}
    \label{fig:lightcurve}
\end{figure*}

\section{X-ray type-I bursts and time-resolved analysis}\label{sec:time-resolved}

AMSPs typically exhibit thermonuclear (type-I) X-ray bursts by unstable burning of accreted matter on the NS surface.
To search for type-I bursts in our observations, we extracted 10\,s-bin light curves from the barycentered \xmm\ EPIC-pn and \nustar\ events.
\xmm\ light curve was then corrected using the \texttt{epiclccorr} task to correct for energy- and time-dependent loss of events.
We found 23 bursts in \nustar\ and 13 in \xmm data. Of these, 9 bursts were observed with both instruments (see Fig.~\ref{fig:lightcurve}). 
We then performed time-resolved spectroscopy of the brightest burst observed simultaneously with \xmm\ EPIC-pn and \nustar\ that, at the same time, happens far from the \nustar\ GTI borders. 
This is burst number 4 in Fig.~\ref{fig:lightcurve}, although we notice here that all bursts throughout the observations considered in this work have similar brightness and recurrence times within $\sim10\%$.
For this, we generated variable-length GTIs according to the burst stage, with shorter GTIs around the peak and longer GTIs during the rise and decay phase.
More specifically, to ensure comparable statistics in all bins, we extracted 4-s length GTIs in the first and last two bins, 3-s GTIs in the third and fourth to last bins, and 2-s GTIs in all others.
We then proceeded to extract source spectra, while we used the persistent source emission spectrum of each instrument obtained in Sect.~\ref{sec:persistent_emission} as a background spectrum. For EPIC-pn spectra, we minimized pile-up effects adopting the same method as described in Sect.~\ref{subsec:xmm_analysis}.

\begin{figure}[!t]
    \centering
    \includegraphics[width=\columnwidth]{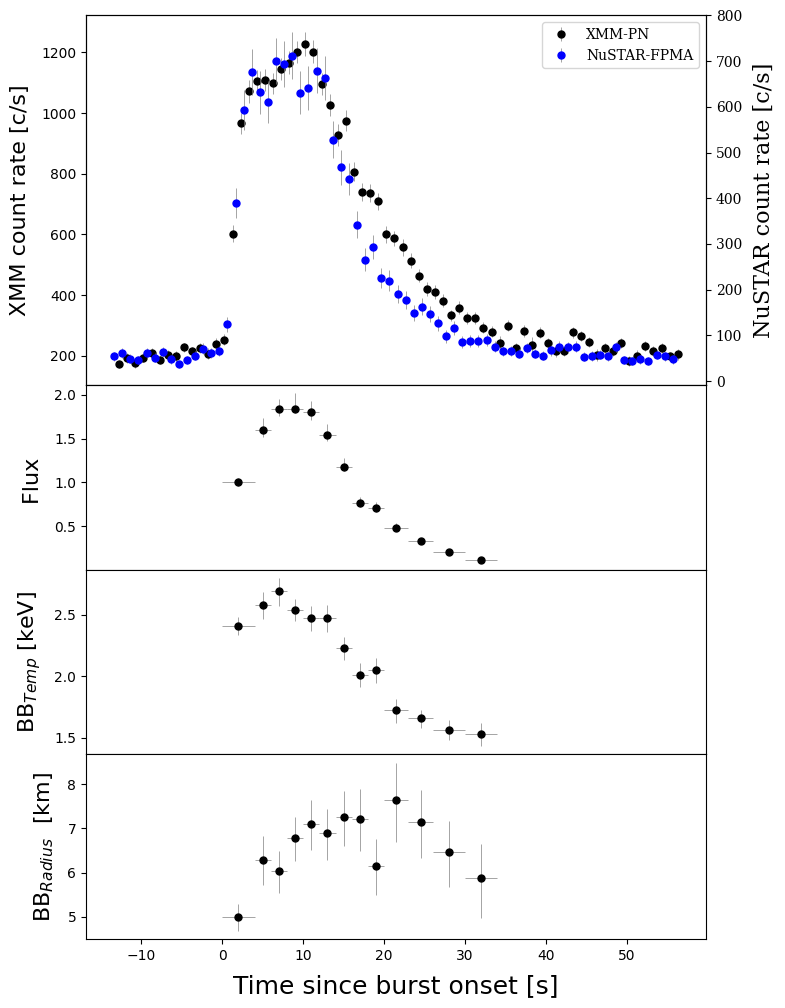}
    \caption{Top panel: light curve of the type-I X-ray burst number~4 (see Fig.~\ref{fig:lightcurve}) observed by EPIC-pn (0.3--10 keV, black dots) and \nustar\ (3--79 keV, blue dots). Unabsorbed flux is traced in the panel below in units of $10^{-8}\,$erg$\,$cm$^{-2}\,$s$^{-1}$ (0.01--15 keV). The burst onset took place at MJD 60368.49178. The temperature and the apparent radius (for a source distance of 8 kpc) of the blackbody component are plotted in the two bottom panels.}
    \label{fig:type1burst}
\end{figure}

Time-resolved spectra over the burst were modeled with an absorbed blackbody component with variable temperature and radius over the burst evolution (see Fig.~\ref{fig:type1burst}), similar to \citet{Molkov2024}.
The persistent emission modeled in Sect.~\ref{subsection:reflection} was employed as the background spectrum. 
This model describes the spectra with reduced $\chi^2\approx1.0$ for all time-resolved bins.
The burst reaches a peak flux of $1.2\times10^{-8}\,$erg\,cm$^{-2}\,$s$^{-1}$ in the 0.5--60~keV band, corresponding to a luminosity of $9.3\times10^{37}\,$erg\,s$^{-1}$ (assuming a distance of 8 kpc).
The blackbody temperature evolves rapidly peaking at about 2.7 keV around the peak of the burst flux and then decreasing steadily. 
The blackbody radius increases  from about 5~km to about 7.5~km around 20~s after the outburst onset (similar to that obtained by \citealt{Molkov2024}).
The data do not show evidence of photospheric radius expansion (PRE). Assuming the empirical Eddington luminosity for hydrogen-poor material of $3.8\times10^{38}\,$erg\,s$^{-1}$ \citep{Kuulkers2003}, we estimated an upper limit on the source distance of about 11.5 kpc (in accordance with the upper limit of 10.6 kpc, \citealt{Ng2024}).

\begin{figure}[!t]
    \centering
    \includegraphics[width=\columnwidth]{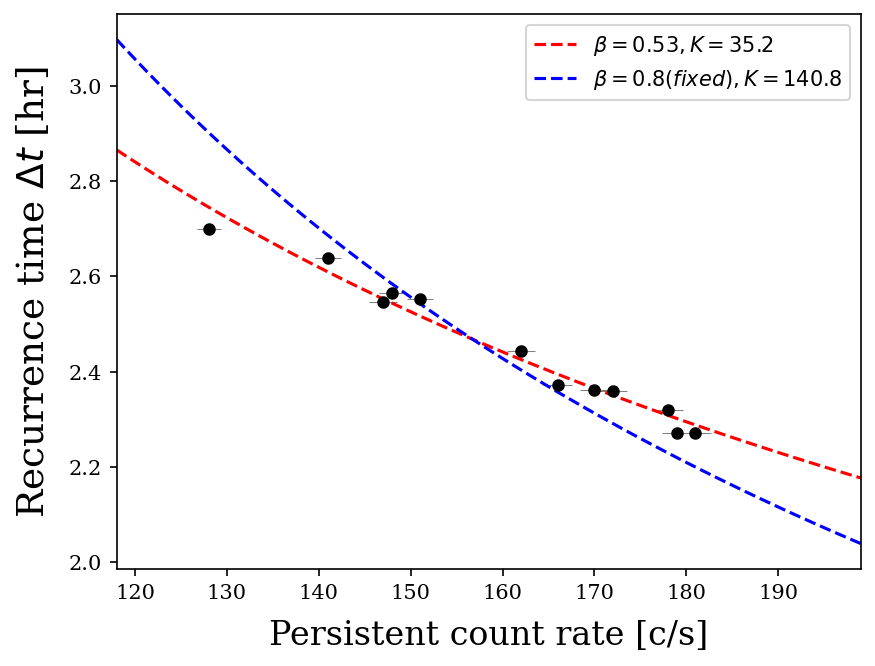}
    \caption{Variation of the recurrence time as a function of the \xmm\ EPIC-pn persistent count rate. The red dashed line represents a fit of data points to Eq.~\eqref{eq:recurrence}. The blue dashed line represents the fit with $\beta$ value fixed to 0.8 (see text).}
    \label{fig:recurrence}
\end{figure}

In the peak of the outburst of \src\ the burst recurrence time $\Delta t$ was increasing with the simultaneously decreasing persistent flux from $\sim1.6\,$hr to $\sim2.9\,$hr \citep{Molkov2024,Sanchez-Fernandez2024a,Sanchez-Fernandez2024b}. Successively, the recurrence time was measured to increase up to $\sim8\,$hr at the end of the \ixpe\ observation \citep{Papitto2025}.
It is typically observed that  
\begin{equation}\label{eq:recurrence}
    \Delta t = K\,C^{-\beta} ,
\end{equation}
where $C$ is the persistent count rate observed right before the burst, $K$ is a proportionality constant, while $\beta$ is the power-law index. Typical measured values of the power-law index are $\beta\sim1$ \citep{Falanga2011, Bagnoli2013}.
Observations of \src\ also have resulted in $\beta\sim$0.8--1.0 \citep{Molkov2024, Papitto2025, Takeda2024}.
With the continuous \xmm\ observation of \src\ we are able to infer the best-fit parameters of Eq.~\eqref{eq:recurrence} without any assumption on the observed quantities ($\Delta t$ and $C$) deriving from data gaps that can sometimes affect  other instruments.
Our results are shown in Fig.~\ref{fig:recurrence}.
Contrary to previous works, we find a power-law index value of $\beta\sim0.5$. We compare our best-fit results to those obtained by \citet{Papitto2025} by fitting our data with Eq.~\eqref{eq:recurrence}. 
For this, in Fig.~\ref{fig:recurrence} we also plot the power-law index derived in their work (fitting only the proportional constant $K$). 

\section{Discussion and conclusions}

We have analyzed \xmm\ and \nustar\ observations of the ASMP \src\ during its 2024  outburst. Our main findings are summarized and discussed in the following sections.

\subsection{Discovery of a relativistic iron line} 

We fitted the persistent emission employing a semi-phenomenological continuum spectral model consisting of a Comptonized component plus two blackbody components, without modeling the reflection component. This model suggests the presence of a relativistically broadened iron emission line, observed for the first time in this source.
When this feature is modeled with a \texttt{diskline} component and all line parameters are let free, the best-fit energy of the line is 6.2 keV.
This is only marginally consistent with neutral iron, contrary to the higher ionization regime implied by the identification of a possible \ion{Fe}{xxvi}   edge in the spectrum.
Nonetheless, low-state iron lines at 6.2 keV have been reported in other systems, such as the black holes Cyg~X-1 and XTE~J1720$-$318 \citep{Barr1985, Markwardt2003}, and in the double pulsar system J0737$-$3039 \citep{Egron2017}.
In any case, by freezing the inclination value parameter of the model used to fit the iron line to the same value obtained by the reflection best-fit model (that is 53\degr) we recover the best-fit energy of the line at 6.4 keV.
We identify this line with the iron K$\alpha$ line of \ion{Fe}{I-XV}.
This is consistent with the relatively low ionization parameter ($\log\xi\simeq2.3$) obtained from the reflection model employed in Sect.~\ref{subsection:reflection}, although it is in contrast with the possible detection of a \ion{Fe}{xxvi} edge discussed more in detail in Sect.~\ref{subsec:discussion_reflection}.
The best-fit parameters of the blurred line component imply a truncated accretion disk at 6\,$R_{\rm g}$ when the inclination of the system is frozen at 53\degr, which is the inclination value obtained by fitting the reflection spectrum (see Sect.~\ref{subsection:reflection}), and employed in order to obtain a best-fit value of the iron line energy that is consistent with the K$\alpha$ iron energy range.
Alternatively, a fit where the system inclination is let as a free parameter returns $85^{+5}_{-14}\,$deg, in broad agreement with polarization measurements of the source which find $74^{+6}_{-6}\,$deg \citep{Papitto2025}, and an inner disk radius of $10.0^{+1.2}_{-1.7}\,R_{\rm g}$.
A possible, qualitative scenario to ease the tension is that of a disk being more inclined in the inner region, where the polarization measurements come from, with the reflection contribution being more relevant from more distant, less inclined parts of the disk.
On the other hand, the discrepancy in the values of the inner disk radius derived from the two fitting procedures is likely a result of the parameters degeneracy. In fact, both inclination and inner disk radius play a major role in shaping the resulting iron line flux and shape \citep{Fabian1989, George1991}, and the observed discrepancy is rather due to the model-driven dependence of the two parameters.

\subsection{The reflection spectrum}\label{subsec:discussion_reflection}

The persistent source emission can also be fitted with a Comptonized spectrum modified by a reflection component deriving from the accretion disk. 
A number of discrete features, both instrumental and physical, are observed in the broadband spectrum, supporting the interpretation of a highly-ionized reflecting medium. The best-fit parameters reveal a moderately hot ($kT_{\rm e}\approx20\,$keV) medium with a thermal component at $kT_{\rm seed}\approx1\,$keV emitted by a region on the NS with $11\,d_{8}\,$km radius (where $d_{8}$ is the source distance in units of 8 kpc), and a hard component interpreted as reflection by the accretion disk at an apparent inner radius of $R_{\rm in}=0.8\sqrt{N_{\rm diskbb} / \cos\,i}\,d_8$ km, that is $27^{+3}_{-3}\,d_{8}\,$km (where $N_{\rm diskbb}$ is the normalization of the \texttt{diskbb} spectral component, and $i=53^\circ$, see Table~\ref{table:spectral_xmm_nustar}).
This is significantly larger than the internal radius value of the accretion disk pegged at the lower limit of 6$R_{\rm g}$ as obtained from best fit of the reflection model, considering that
the definition of the gravitational radius adopted in the model is $R_{\rm g}=GM_{\rm NS}/c^2\simeq2\,$km for a NS mass of $1.4\,{M_\odot}$.
However, such inconsistency between the two methods of retrieving the inner disk radius is often observed throughout the literature (see, e.g., \citealt{Cackett2010,Ludlam2017}). This is typically attributed to possible boundary layer corrections and other factors \citep{Kubota2001}, including the scattering of photons off the line of sight by the accretion disk, with the net effect of reducing the contribution of the blackbody component of the disk and thus the inferred apparent radius.
We also notice that the relatively low plasma temperature ($kT_{\rm e}\simeq20\,$keV) observed here is consistent with polarimetric predictions by \citet{Bobrikova2023}, as measured by \citet{Papitto2025}.

The reflected component implies a sub-Solar iron abundance (despite a prominent iron emission line, see also later), with a blurring convolution that returns an inclination of 53\degr.
This is in agreement with previous results obtained by modeling of the pulsed emission, which gives an inclination of $58\degr$ and an internal disk radius of $24.6\,$km  \citep{Molkov2024}, for which they estimate an error of less than $10\%$, but significantly deviating from the value obtained by modeling of the phase-resolved polarized emission which returned an inclination of $74^{+6}_{-6}\,$deg \citep{Papitto2025}.
Moreover, assuming that the emitting region distance is coincident with the inner disk radius, we can use the definition of $\xi$ to infer the inner disk radius, that is $\xi=L_{\rm X}/(n_{\rm H}\,R_{\rm in}^2)$ \citep{Tarter1969}.
The bolometric $L_{\rm X}$ luminosity can be obtained by the bolometric flux reported in Table~\ref{table:spectral_xmm_nustar} as $2.5\times10^{37}\,$erg\,s$^{-1}$ at a distance of 8 kpc.
A typical Shakura-Sunyaev disk has density $n_{\rm H}\sim5\times10^{20}\,$cm$^{-3}$ (see, e.g., \citealt{Ross2007}).
This results in an inner radius of about $150\,$km, much larger than the value estimated above. 
This difference can be mitigated considering that the luminosity irradiating the disk is inferior to bolometric luminosity by a geometry-dependent factor $\zeta$ \citep{Zdziarski2020},
and assuming a slightly higher disk density, e.g., $n_{\rm H}=6\times10^{20}\,$cm$^{-3}$, which results in an inner radius of $40\pm4\,$km.
However, we notice that the inner disk radius for a 447.9 Hz pulsar cannot be significantly larger than the co-rotation radius of 28.6$\sqrt{M_{\rm NS}/1.4M_\odot}\,$km. 

The fit also requires a broad gaussian emission line at a centroid energy of about 3.3 keV, which we identify with \ion{Ar}{xviii}. Its large width is likely due to relativistic smearing, as supported by the iron emission line shape (see Sect.~\ref{subsection:iron_line}). However, the reflection component should already account for emission lines from ionized elements, such as the \ion{Ar}{xviii}, but also those corresponding to typically more abundant elements such as \ion{Si}{xiv} and \ion{S}{xvi}.
The fact that our best-fit model requires an additional gaussian emission component only for the \ion{Ar}{xviii} is therefore unexpected.
We interpret this as probably due to the relatively low best-fit ionization parameter, which makes the \ion{Ar}\,line computed from the reflection component alone not intrinsically strong enough to account for the actual observed line, thus requiring an additional component.
We also notice the peculiar absorption edge at 9.68 keV.
Nominally, this energy value would be consistent with a Zn edge at 9.65 keV.
However, the low abundance of Zn makes it highly unlikely.
On the other hand, the presence of \ion{Ar}{xviii} argues for highly ionized elements and is therefore in agreement with the presence of \ion{Fe}{xxvi}.
If the feature actually originates from \ion{Fe}{xxvi}, the difference between the nominal (9.278 keV) and the observed (9.68 keV) energy can be interpreted in terms of a blueshifted wind feature.
Similar features characteristic of an outflowing wind have been observed in other AMSPs as well \citep{Degenaar2017, Allen2018, Nowak2019}, and in other high-inclination systems (see, e.g., \citealt{Reeves2004}).
The resulting outflow speed of ${\sim}13,000\,$km\,s$^{-1}$, that is $\sim$4\% the speed of light, would place such a wind in the regime of ultrafast outflows \citep{Miller2015, Miller2016a}.
This is in agreement with the observed low ionization parameter of the reflecting medium, which facilitates radiation pressure acceleration, although it clashes with the very interpretation of a fully ionized iron state. The low iron abundance derived by the reflection model may resolve this tension if iron has a sub-solar abundance, but other elements do not.
This is especially interesting given the preliminary evidence for the non-solar elemental composition in \src\ \citep{Dohi24}.
In particular, \citet{Dohi24} provide evidence of high $Z_{\rm CNO}$ abundances compared to solar values. 
Carbon-enhanced metal poor (CEMP) stars are a distinct population of stars in their main sequence or giant phase, often found in binary systems \citep{Cooke2014, Hirai2025}.
This is consistent with the M-type main-sequence companion star suggested for \src\ \citep{Ng2024}.
If confirmed, this would be the first time a CEMP star is observed feeding a compact object.  

\subsection{Burst energetics}

The broadband time-resolved analysis of type-I bursts does not show evidence of photospheric expansion. The recurrence time $\Delta t$, which can be accurately estimated thanks to \xmm\ continuous observation, reveals a dependence on the persistent count rate that is milder than previously observed.
We ascribe the observed difference to different factors. First, the narrow interval of $\Delta t$  covered by our observations can affect the best-fit results. Second, we notice the hysteresis in Fig.~10 of \citet{Papitto2025}, where similar count rate values exhibit $\Delta t$ values that are sensibly different (up to ${\sim}1\,$hr), a condition that can heavily drive the fit results, especially given the above-described factor. Third, there might be a physical reason behind the power-law index difference due to the different energy bands explored and a possible evolution of the spectrum over the outburst, although no clear evidence for this has been observed so far \citep{Gilfanov1998, Gierlinski2005, Poutanen2006}.

By combining the observed parameters of persistent and burst emission, our data set allows us to obtain the $\alpha$-value (see \citealt{Galloway2022} for the definitions employed below), that is, the ratio of accretion to thermonuclear burning energy:

\begin{equation}\label{eq:alpha1}
    \alpha = \frac{\Delta t\, F_{\rm pers}}{E_{\rm burst}} ,
\end{equation}
where $F_{\rm pers}$ is the bolometric flux of the persistent emission, $E_{\rm burst}$ is the burst fluence (i.e., the burst flux integrated over the burst time), and $\Delta t$ is the burst recurrence time.
This is important to evaluate the composition of the bursting fuel in terms of the mean hydrogen mass fraction at ignition $\bar{X}$.
For burst number 4 (see Fig.~\ref{fig:lightcurve}), the total unabsorbed flux (0.01--100 keV) is $\sim1.5\times10^{-8}\,$erg\,cm$^{-2}\,$s$^{-1}$, which gives a burst fluence  of $E_{\rm burst}=3.1^{+0.2}_{-0.2}\times10^{-7}\,$erg\,cm$^{-2}$, calculated from the time-resolved spectroscopy fluxes (see Sect.~\ref{sec:time-resolved}).
We therefore estimate $\alpha=76^{+3}_{-3}$ for burst number 4.
In comparison, \citet{Molkov2024} derive $\alpha\simeq105$, although at a different outburst stage, when the composition of the fuel was likely different. 
On the other hand, the $\alpha$-value can also be calculated as: 
\begin{equation}\label{eq:alpha2}
    \alpha = \frac{Q_{\rm grav}}{Q_{\rm nucl}}\frac{\xi_{\rm b}}{\xi_{\rm p}}(1+z) ,
\end{equation}
where $\xi_{\rm b}$ and $\xi_{\rm p}$ are the burst and persistent anisotropy factors accounting for any beaming effects. $Q_{\rm grav}$ and $Q_{\rm nucl}$ are the accretion and burst energy generation per unit mass, respectively. Following \citet{Galloway2022}, $Q_{\rm grav}=c^2z(1+z)\approx GM_{\rm NS}/R_{\rm NS}$, with $z$ being gravitational redshift that, for a typical NS of mass $M_{\rm NS}=1.4\,M_\odot$ and radius $R_{\rm NS}=11.2\,$km, is $z=0.259$. Thus $Q_{\rm grav}\approx190\,$MeV\,nucleon$^{-1}$. On the other hand, the exact value of $Q_{\rm nucl}$ depends on the fuel composition and, for a mixed H/He layer, the approximate expression $Q_{\rm nucl}=(1.31 + 6.95\bar{X}-1.92\bar{X}^2)\,$MeV\,nucleon$^{-1}$  has been recently adopted \citep{Goodwin2019}.
Provided that the $\alpha$-value is known by observational means as done above through Eq.~\eqref{eq:alpha1}, then 
by substituting the approximate expression for $Q_{\rm nucl}$ in Eq.~\eqref{eq:alpha2} one can obtain \citep{Galloway2022}:
\begin{equation}\label{eq:mean_H}
    \bar{X} = z\frac{155}{\alpha}\frac{\xi_{\rm b}}{\xi_{\rm p}}-0.23 .
\end{equation}

Assuming $\xi_{\rm b}$/$\xi_{\rm p}$ of the order of unity, we thus obtain $\bar{X}=0.29(3)$. This would imply a heavily H depleted fuel (with respect to solar abundances $\bar{X} =0.7$). 
The resulting burst energy generation is therefore $Q_{\rm nucl}=3.16\,$
MeV nucleon$^{-1}$.
While the burst is likely burning a mix of H/He, during the persistent emission steady H-burning via hot-CNO cycle is happening. The time scale for this is (10--15~hr)$(X_0/0.7)(Z_{\rm CNO}/0.02)^{-1}(1+z)$, where $X_0$ is the fraction of the accreted hydrogen mass \citep{Lampe+2016}. This is significantly longer than the burst observed recurrence time, consistent with observations of He bursts taking place in a medium where H has not been completely depleted. However, the H abundance we observed ($X=0.3$) is much lower than what would be expected after a $\geq2$ hours recurrence time. This might hint at a subsolar H fraction of the accreted material. 

We can also estimate the burst ignition depth (see \citealt{Galloway2022}, Eq.~4):
\begin{equation}\label{eq:y_ign}
    y_{\rm  ign} = \frac{L_{\rm  burst}(1+z)}{4\pi R_{\rm  NS}^2Q_{\rm  nucl}} , 
\end{equation}
obtaining a value of $y_{\rm  ign}=0.17\,\xi_{\rm  b}\, d_8^2\, 10^{8}\,$g\,cm$^{-2}$. 
We can then evaluate the expected value of the recurrence time as $\Delta t = y_{\rm  ign}(1+z)/\dot{m}$, where $\dot{m}=L_{\rm  pers}(1+z)/[4\pi R_{\rm  NS}^2(GM_{\rm  NS}/R_{\rm  NS})]$ is the local accretion rate. We obtain $\dot{m}= 7.7\times10^{3}\xi_{\rm  p}\,d^{2}_8\,$g\,s$^{-1}$ which, in turn, returns an expected recurrence time of $\Delta t = 2765\,$s. 
This is about a factor of three smaller than the observed average value of 2.4 hr. A similar discrepancy between the expected and observed value of the recurrence time was also obtained by \citet{Falanga2012} for the AMSP system IGR~J17498$-$2921, attributed to peculiar fuel composition (decreased CNO or $X_0$) combined with an overestimation of the isotropic luminosity.
An even lower H fraction than indicated by the burst energetics would increase the burst energy generation rate and the burst ignition depth, so leading to a longer expected burst recurrence rate. If the bursts were taking place in a pure He environment (i.e., $X=0$), the expected recurrence time at the observed count rate would amount to $\sim1.85$ hr, much closer to the observed value. Alternatively, such a discrepancy can be mitigated if the persistent emission is beamed towards the observer (that is, $\xi_{\rm p
}<0.5$),
or if a larger gravitational redshift is taken into account due to a massive NS as suggested for \src\ \citep{Takeda2024}.


\begin{acknowledgements}
C.M., A.P., G.I., and R.L.P are supported by INAF (Research Grant `Uncovering the optical beat of the fastest magnetised neutron stars 620
(FANS)') and the Italian Ministry of University and Research (MUR) (PRIN 2020, Grant 2020BRP57Z, `Gravitational and Electromagnetic-wave Sources in the Universe with current and next-generation detectors (GEMS)'). A.P. acknowledges support from the Fondazione Cariplo/Cassa Depositi e Prestiti, grant no. 2023-2560.
J.P. was supported by the Ministry of Science and Higher Education grant 075-15-2024-647. ADM contribution is supported by the Istituto Nazionale di Astrofisica (INAF) and the Italian Space Agency (Agenzia Spaziale Italiana, ASI) through contract ASI-INAF-2022-19-HH.0.
\end{acknowledgements}

%
%

\bibliographystyle{yahapj}
\bibliography{references}

\end{document}